\documentclass[fleqn,10pt]{wlscirep}
\usepackage[utf8]{inputenc}
\usepackage[T1]{fontenc}
\title{Temporal Evolution of Risk Behavior in a Disease Spread Simulation}

\author[1,2,3,4,*]{Ollin D. Langle-Chimal}
\author[3,4,5]{Scott C. Merrill}
\author[3,5]{Eric M. Clark}
\author[3,5]{Gabriela Bucini}
\author[3,4]{Tung-Lin Liu}
\author[3,7]{Trisha R. Shrum}
\author[3,8]{Christopher Koliba}
\author[1,3,4,8]{Asim Zia}
\author[3,6]{Julia M. Smith}
\author[1,2,3,4]{Nicholas Cheney}

\affil[1]{Department of Computer Science, The University of Vermont, Burlington, VT, USA}
\affil[2]{Complex Systems Center, The University of Vermont, Burlington, VT, USA}
\affil[3]{Social Ecological Gaming and Simulation Lab, The University of Vermont, Burlington, VT, USA}
\affil[4]{Gund Institute for Environment, The University of Vermont, Burlington, VT, USA}
\affil[5]{Department of Plant and Soil Science, The University of Vermont, Burlington, VT, USA}
\affil[6]{Department of Animal and Veterinary Sciences, The University of Vermont, Burlington, VT, USA}
\affil[7]{Department of Community Development and Applied Economics, The University of Vermont, Burlington, VT, USA}
\affil[8]{School of Public Affairs and Administration, The University of Kansas, Lawrence, KS, USA}

\affil[*]{olangle@uvm.edu}

\begin{abstract}
Human behavior is a dynamic process that evolves with experience. Understanding the evolution of individual's risk propensity is critical to design public health interventions to propitiate the adoption of better biosecurity protocols and thus, prevent the transmission of an infectious disease. Using an experimental game that simulates the spread of a disease in a network of porcine farms, we measure how learning from experience affects the risk aversion of over $1000$ players. We used a fully automated approach to segment the players into four categories based on the temporal trends of their game plays and compare their overall game performance. We found that the risk tolerant group is $50\%$ more likely to incur an infection than the risk averse one.  We also find that while all individuals decrease the amount of time it takes to make decisions as they become more experienced at the game, we find a group of players with constant decision strategies who rapidly decrease their time to make a decision and a second context-aware decision group that contemplates longer before decisions while presumably performing a real-time risk assessment. The behavioral strategies employed by players in this simulated setting could be used in the future as an early warning signal to identify undesirable biosecurity-related risk aversion preferences, or changes in behavior, which may allow for targeted interventions to help mitigate them. 
\end{abstract}
\begin{document}

\flushbottom
\maketitle
\thispagestyle{empty}

\section*{Introduction}

Recent events have shown the importance of implementing biosecurity measures to combat infectious diseases. Biosecurity measures take various forms and are context-specific, ranging from simple Non-Pharmaceutical Interventions (NPIs) like mask-wearing and social distancing for human viral diseases such as COVID-19~\cite{flaxman2020estimating}, to more complex processes like comprehensive sanitation and flow control in farm facilities~\cite{alarcon2021biosecurity}. The COVID-19 pandemic has demonstrated that individual adoption of prophylactic measures is heterogeneous across the population~\cite{allen2022predicting}, and accounting for this heterogeneity is crucial in designing effective interventions~\cite{twahirwa2020covid}.

Many highly infectious diseases in humans in recent years have been zoonotic in origin, including COVID-19~\cite{contini2020novel}, monkeypox~\cite{parker2007human}, swine flu (H1N1)~\cite{gatherer20092009}, avian flu (H5N1)~\cite{kandeel2010zoonotic}, and even HIV~\cite{hemelaar2012origin}.
This highlights the crucial role of farms as potential hotspots for the emergence of new zoonotic diseases. Additionally, animal diseases in the farm industry, such as Porcine Epidemic Diarrhea Virus (PEDv)~\cite{jung2020porcine} and Swine Fever~\cite{edwards2000classical}, not only threaten animal well-being, but can also severely disrupt food supply chains, leading to substantial economic impacts~\cite{pitts2019impact}, and compromising food security for individuals~\cite{hashem2020animal}.

Given the high health risks, it is crucial to study human behavior in the context of biosecurity propensity when exposed to contagious diseases. Identifying populations with the highest and lowest risk of not adopting health measures enables the implementation of targeted interventions, optimizing resources and outcomes. Experimental approaches in public health risk situations are ethically and financially infeasible, prompting the increased use of serious games in recent years~\cite{shapley} for applications like environmental economics~\cite{haurie}, natural resource management~\cite{Li}, and decision-making~\cite{kroll}. Serious games have been employed to quantify various aspects of human behavior, from assessing moral decisions involving others' lives~\cite{awad2018moral} to understanding generosity propensity~\cite{lotti2022generosity}, promoting healthier behaviors~\cite{jakicic2020gamification}, and increasing empathy and social compliance~\cite{BAILEY2019101052, shrum2021pro}. In the context of the current political polarization of public health policies, it is understandable to consider inferring behavior from demographic data. However, previous studies~\cite{yang2022identifying}, have demonstrated that certain behaviors cannot be characterized or explained based on demographics. Our study aligns with these findings, as we observed no significant demographic differences in game performance or risk behavior (Supplementary Material 1). These results highlight the crucial role of behavioral data in understanding and analyzing human behavior accurately.

Previous research has shown that message format and content influence risk perception and subsequent decision-making~\cite{merrill2019decision}. Contextual factors like information accessibility, content, and format trigger distinct behavioral responses~\cite{clark2019using}, suggesting that effective risk communication is vital for promoting biosecurity adoption during an outbreak. This reinforces the notion that community engagement is crucial in fostering behavior change to slow disease spread~\cite{Bedson661959, skrip2020unmet}. Our study builds upon this premise by investigating the temporal aspect of decision-making in a serious game, characterizing risk aversion across each round played by players (n=1095) in a game simulating PEDv spread in a network of porcine farms~\cite{scott}. The participant pool includes 50 farm professionals recruited at the 2018 World Pork Expo and 1,045 players from Amazon Mechanical Turk. The overall risk propensity between these groups shows no significant difference, as indicated by Clark et al.~\cite{clark2021emulating}, and thus we treat them as one group.

The primary objective of this study was to develop an automated method for discerning distinct trajectories in game strategies throughout the game's entirety and comparing the outcomes generated by each of them. Upon identifying these strategies, we aimed to uncover varying behavioral responses among individuals. These differences can be used as early warnings for shifts in an individual's risk response. These early warnings can be in the form of the time taken by the players to make a decision. For example, those players that exhibit adaptive behavior, exemplified by the so-called \textit{Opportunist} described by Clark et al.~\cite{clark2019using}, tend to be more analytical and, on average, require longer durations to make a move compared to those maintaining a consistent strategy. By recognizing these patterns, we can detect users at risk of a behavioral change and leverage this information to make targeted interventions.

\section*{Results}

\subsection*{Identifying Similarity and Differences in Behavior}\label{identifymethods}

In this study, we analyzed each round of a serious game in which the player is in charge of managing one of 50 farms within a constrained area, over 32 rounds. All other farms are computer controlled. In each round, the player may attempt to prevent their farm from being infected by increasing their farm's biosecurity levels in 3 incremental steps. At each decision point players must weigh the trade-offs between the investment costs of their chosen biosecurity measures and the associated mitigation of infection risk those measures provide. At each round, players are presented with different uncertainty levels of multiple variables, such as contagion rates and biosecurity levels of other farms. A round ends after 6 decision points or upon infection.  In the case of infection a significant penalty is applied to the infected farm's game score (and thus the game-score-proportional monetary payout to the player to promote engagement). A more detailed description of the game, can be found in the methods section. The game was played by 1,095 players. In order to analyze their behavior, they were characterized using a risk aversion metric ($\rho$). This metric represents the propensity of players to take risks by measuring their likelihood of investing in an additional biosecurity level when given the opportunity.

\begin{figure}[!hbp]
    \centering
    \includegraphics[width=0.8\textwidth]{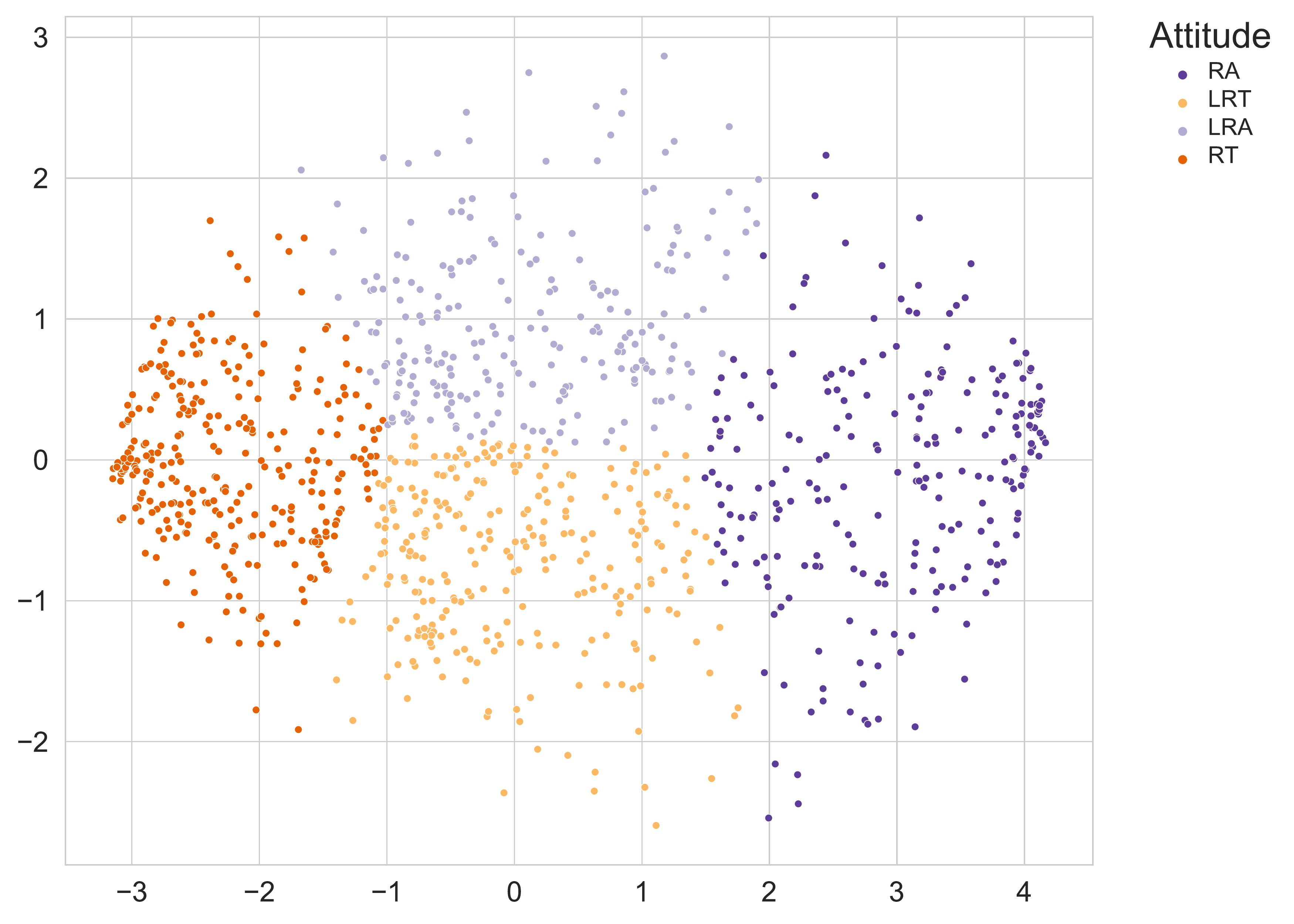}
    \caption{First 2 principal components of the Isomap projection of the vectors composed by the 32 $\rho_r^p$ per player. The colors and group names [Risk Averse (RA), Risk Tolerant (RT), Learning to be Risk Averse (LRA) and Learning to be Risk Tolerant (LRT)] are determined by the temporal behavioral response to biosecurity investment using 4 groups clustering via K-means.}
    \label{fig:allquadrant}
\end{figure}

The 32 randomized treatments rounds allow the representation of each player's entire gameplay through a 32-dimensional vector, denoted as $\vec{\rho^p} = [\rho^p_1, \rho^p_2, \hdots, \rho^p_{32}]$. Each dimension corresponds to the individual $\rho_r$ at round $r$ for player $p$. Using these vector characterizations it was found that it is possible to detect similarities among players while incorporating temporal information.  The vectors were projected into a two-dimensional space using the Isomap algorithm~\cite{balasubramanian2002isomap} to make the data less sparse~\cite{fodor2002}.  Clustering on the resulting representations via K-means~\cite{kmeans, bellman1961}, it was determined (via the distortion metric) that four distinct behavioral clusters were present in this space (Fig.~\ref{fig:allquadrant}).

Interpreting the behaviors within each of these clusters, it was found that two clusters exhibited relatively stable risk aversion metrics throughout the game. The first cluster(n=251) had a median $\rho \sim 1$, meaning that these players consistently maintained the maximum possible biosecurity investments, and we thus label them as \textit{Risk Averse} players (RA). In contrast, the second cluster (n=337), with a median $\rho \sim 0$ representing virtually no biosecurity investment, was designated \textit{Risk Tolerant} (RT). 

\begin{figure}[!hbp]
    \centering
    \includegraphics[width=0.8\textwidth]{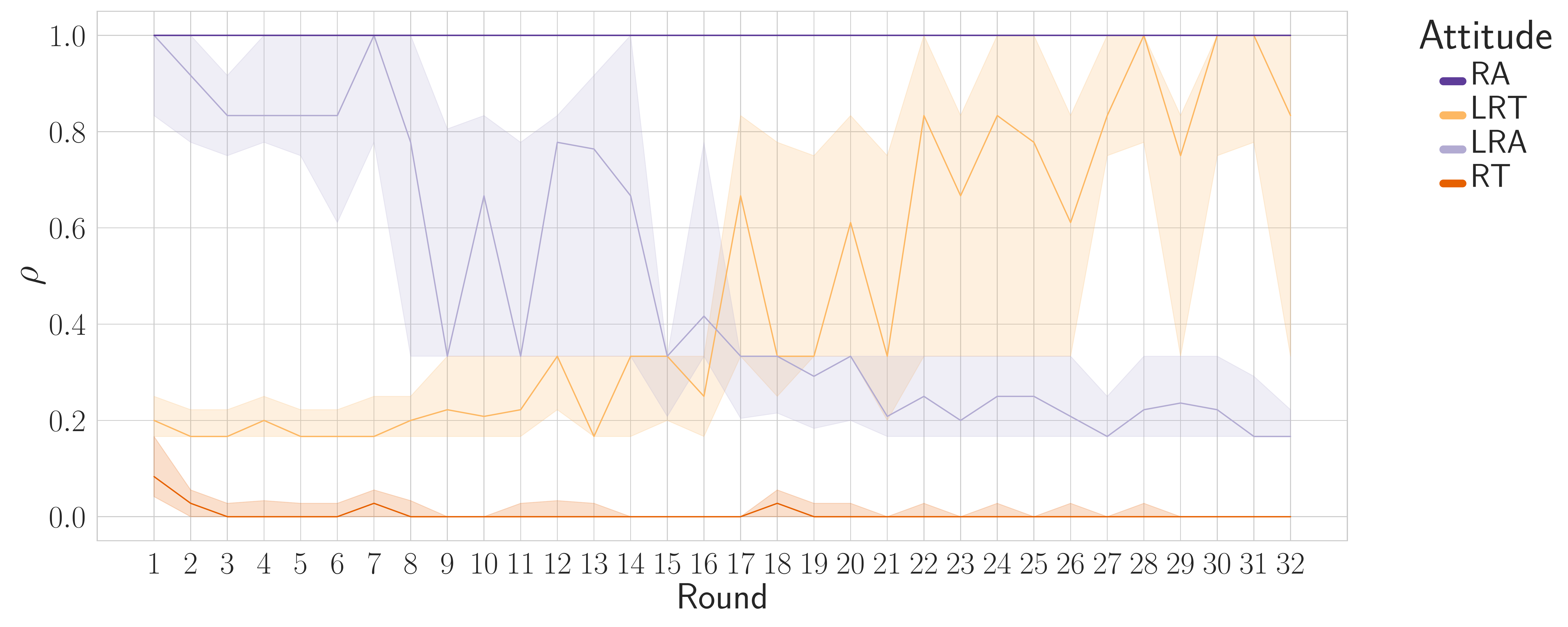}
    \caption{Median $\rho$ by group and by round with 95\% CI bounds. Four principal behavior patterns are found by this approach. The first two (RA and RT) maintain relatively the same levels of $\rho$ at each round while the remaining ones first show a convergence to later divert again. One of those groups starts the game with a high biosecurity investment ratio and gradually decreases it, the other group shows the opposite behavior. By these patterns we named our groups as \textit{Risk Tolerant} (RT) if their median investment is $0$, \textit{Risk Averse} (RA) if their median $\rho=1$, the group with an increasing investment will be \textit{Learning to be Risk Averse} (LRA) and the decreasing one \textit{Learning to be Risk Tolerant} (LRT).}
    \label{fig:medianpatterns}
\end{figure}

The remaining two clusters both showed an intermediate level of biosecurity investment, and the distribution of $\rho$ values was not significantly distinguishable between the two groups (Mann-Whitney, $U=26861$, $p=0.056$, two-sided).  But despite the similar aggregate amount of investment in biosecurity, visual inspection of the two groups in Fig.~\ref{fig:medianpatterns} shows a clear differentiating factor: shifts in behavioral preferences over time.  One group (n=279) displayed increasing $\rho$ values, where increased experience in the game lead to an increase in biosecurity investments, and was named Learning to be \textit{Risk Averse} (LRA).  The other group (n=228) showed decreasing investments when gaining gameplay experience and was called Learning to be \textit{Risk Tolerant} (LRT). Notably, all four of above groups were of approximately equal size, see Table~\ref{table:time} for reference.

% We found that when considering only aggregated measures rather than round-dependent ones, there aren't significant differences between the learning groups. 
The Session Profit represents the player's final score and monetary outcome after considering the cost of biosecurity investments and monetary risks associated with disease outbreaks (both within the game and corresponding with real payout amounts to participants in an attempt to incentivize profit maximization during gameplay). Interestingly there wasn't any significant difference in Session Profit between any of the four groups (all $p \geq 0.51$), suggesting that no one strategy is objectively better or worse than another in this game or necessarily determined by quality of the player.
The largest Session Profit difference was found between the LRT and the RA groups (Mann-Whitney, $U=38444$, $p=0.5141$) with median values of \$154,000 and \$149,000 respectively. 
Regarding the $\rho$ measure, all groups show significant differences with the smallest one being from the two learning groups (Mann-Whitney, $U=26861$, $p=0.0024$, two-sided). For the number of infections by group, the only non significant difference was between learning groups (Mann-Whitney, $U=31203$, $p=0.7102$, two-sided) with median values of  5 for both of them, in contrast, it was 6 for the RT,and 4 for the RA.

\begin{figure}[!hbp]
\centering
\includegraphics[width=\textwidth]{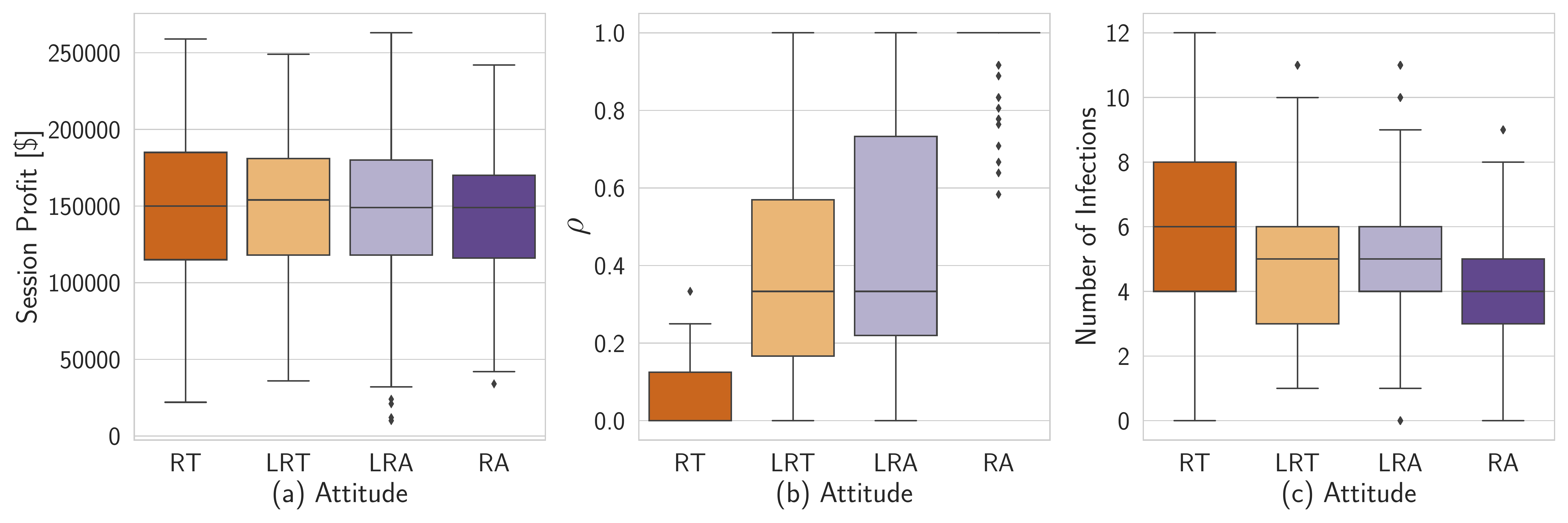}
\caption{
The learning groups cannot be discern from each other using the aggregated collected game variables such as (a) final session profit, (b) the mean risk aversion metric nor (c) the number of infection occurrences.
}
\label{fig:measures_attitudes}
\end{figure}

It is important to note that the RT group's expected number of infections was significantly higher than that of the RA group (Mann-Whitney, $U=63706$, $p < 0.001$, two-sided), as shown in Figure~\ref{fig:measures_attitudes}(c). This result indicates that systematically increasing biosecurity adoption substantially impacts infection occurrences. However, the final game score, or \textit{Session Profit}, is not significantly affected by this difference Figure~\ref{fig:measures_attitudes}(a), meaning that the mean cost of infection for the RA group is higher than that for the RT group, balancing out, on average, the additional cost of biosecurity investments from risk averse players.

% In contrast to the aggregated values of the entire gameplay, the median values for each group at every round display an inverse behavior between the LRA and LRT groups. The LRA group begins the session with low $\rho$ values and later shifts to higher values, while the LRT group does the opposite, starting with high values and later decreasing them. In both cases, the variance is lower when the mean value is in the lower bound than in the upper bounds, where the median even exhibits a less smooth behavior. As expected, the RA and RT groups are characterized by steady lines at the extreme values of 0 and 1, with the 95\% CI bounds only visible in three rounds for the RT group. In the following rounds, they are indistinguishable.

%%%%%%%%%%%%%%%%%%%%%%%%%%%%%%%%%%%%%%%%%%%%%%%%%%%%%%%%%%%%%%%
%%%%%%%%%%%%%%%%%%%%%%%%%%%%%%%%%%%%%%%%%%%%%%%%%%%%%%%%%%%%%%%
%%%%%%%%%%%%%%%%%%%%%%%%%%%%%%%%%%%%%%%%%%%%%%%%%%%%%%%%%%%%%%%
%%%%%%%%%%%%%%%%%%%%%%%%%%%%%%%%%%%%%%%%%%%%%%%%%%%%%%%%%%%%%%%
%%%%%%%%%%%%%%%%%%%%%%%%%%%%%%%%%%%%%%%%%%%%%%%%%%%%%%%%%%%%%%%
%%%%%%%%%%%%%%%%%%%%%%%%%%%%%%%%%%%%%%%%%%%%%%%%%%%%%%%%%%%%%%%
%%%%%%%%%%%%%%%%%%%%%%%%%%%%%%%%%%%%%%%%%%%%%%%%%%%%%%%%%%%%%%%
%%%%%%%%%%%%%%%%%%%%%%%%%%%%%%%%%%%%%%%%%%%%%%%%%%%%%%%%%%%%%%%

% If we look at the scaled density distribution of the $\rho$ values for each group we can't find a significant difference between the LRA and LRT groups (Mann-Whitney, $p=0.056$, two-sided). Which suggest that the aggregation of these values by group fails to capture the dynamic nature of behaviors.

\begin{figure}[!hbp]
\centering
\includegraphics[width=\textwidth]{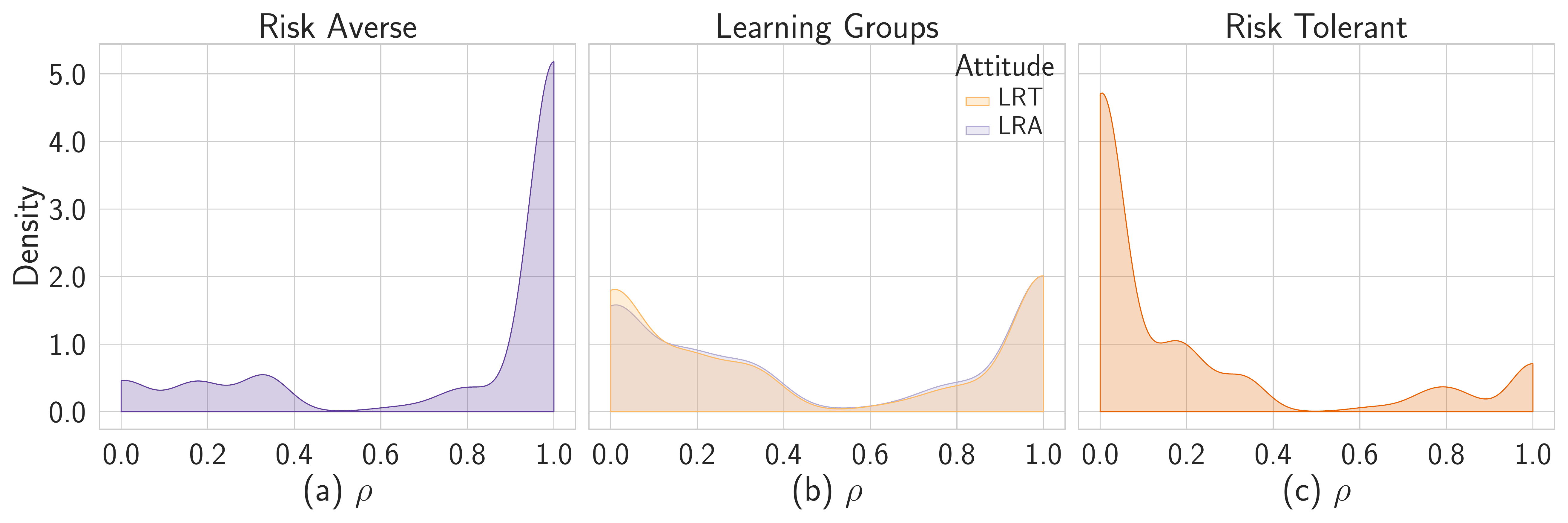}
\label{fig:density1}
\caption{
Probability density of the proportion of investment in biosecurity by group. These clusters are only distinguishable if we consider behavior over time, and are not present in we look at aggregated values of gameplay statistics.
}
\end{figure}

Characterizing the evolution of the risk aversion of the players shows promising insights to a better understanding of the diverse behavioral responses that the players exhibited when presented with a risk of infection. Specifically, it was demonstrated the possibility to detect temporal shifts in behavioral preferences, evidenced by the two Learning groups. This provides valuable insights into how players react to risk in a dynamic and uncertain environment, and the complexities of optimal strategies in such settings.

\subsection*{Time to make a play}\label{time}

The study of the time taken by each group to complete a game round serves as a metric for detecting early warning that could signal a behavioral change. Our cognitive processes are intricately linked with time, and decision-making speed can offer an insightful perspective into the strategic considerations and adaptive behaviors of players~\cite{payne1988adaptive,hinson2003impulsive}.
Within the context of this game, longer decision times can suggest meticulous analysis, risk evaluation, and strategy formulation. Thus, time analysis may provide a reliable early warning signal for a shift towards more responsive game strategies.
Understanding these shifts is crucial as they could reflect broader patterns of human behavior in response to uncertainty and risk. This could be a pivotal component for designing targeted interventions in a broad set of contexts, such as behavioral economics~\cite{kahneman2011thinking}, risk analysis~\cite{slovic2004risk}, and more relevant to this study, health policy planning. To the best of our knowledge, this is the first study to systematically explore the relationship between the time taken to make a decision and the assessment of biosecurity risk.

\begin{figure}[!hbp]
    \centering
    \includegraphics[width=0.9\textwidth]{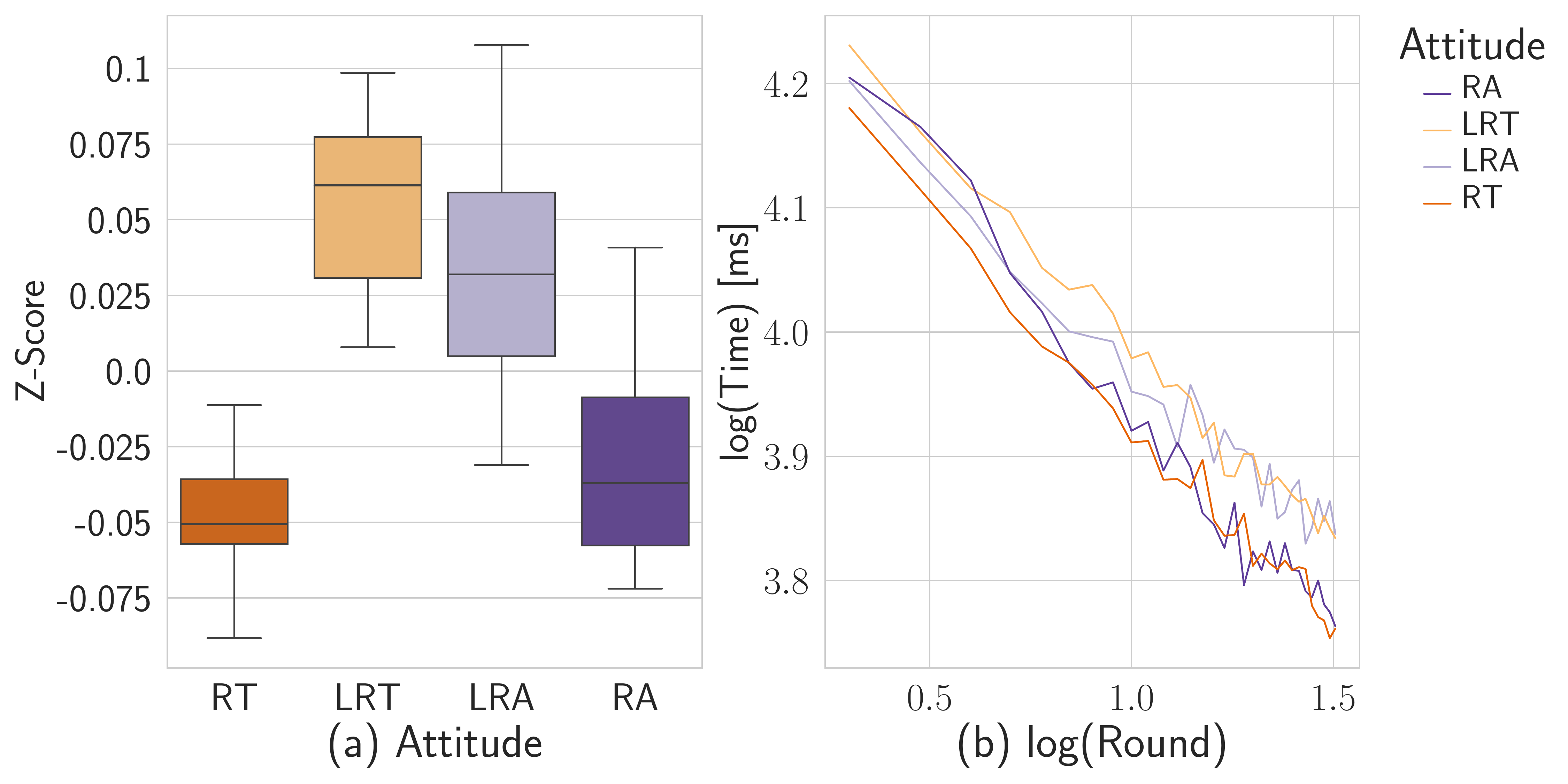}
    \caption{Individuals who change their behavior take more time, on average, to make decisions.  We also find that in the individuals who play with a non-learning behavior, those that are risk tolerant spend significantly less time making decisions than those two are risk averse. (a) Median Z-Score of the time spent per round. (b) Median time spent per round by behavioral group. Taking out the first round which took significantly a longer time to complete for all groups, we found that the time spent per round decayed as a power-law for all of the groups.}
    \label{fig:timeall}
\end{figure}

To compare the average round completion time among groups, we calculated the withing-group median z-score of the time elapsed before the player took an action, stratified by random treatment type. Figure~\ref{fig:timeall}(a) displays the distribution of median values for these differences, aggregated by round.
Considering that different treatments are presented in random order to each player, we disregard their average effect on time. On average, players exhibiting adaptive behavior take longer to play compared to those employing a steady strategy. The smallest difference between groups lies between the two non-learning groups (Mann-Whitney, $U = 139$, $p = 0.0314$, two-sided), with the absolute value of the median z-score for the RT being 36.5\% times larger than that of the RA. The most significant difference is observed between the RT and LRT groups (Mann-Whitney, $U = 0$, $p = 3.2e-12$, two-sided), with the median z-score of the LRT group 221.4\% larger than that of the RT. This finding supports our hypothesis that players with adaptive behavior, on average, require more time to make decisions in contrast to those utilizing a consistent strategy.

These results suggest that behavioral adaptation may demand additional time for individuals to contemplate alternative choices or that individuals predisposed to adapt need to process supplementary information from their surroundings to make well-informed decisions. Figure~\ref{fig:timeall}(b) delineates the progression of median time spent at each round minus the first, on a log-log scale, revealing a power-law decay $[f(t) = at^{-k}]$ across all groups. We removed the first round as players consistently spent more time on it while they get comfortable with the game. It is particularly striking that, during the final rounds of the game, the learning groups converge, as do their non-learner counterparts, while both meta-groups demonstrate divergence between each other. The RA group shows the fastest decay with a $k = 0.3650$, which is 29.89\% larger than the slowest decay exhibited by the LRA group with $k = 0.2810$.

This observation implies that as the game progresses, individuals displaying adaptive behavior start to reveal similarities in their decision-making time, while those adhering to a consistent strategy exhibit divergence from the former. The accelerated decay of the RA group might suggest that, over time, these individuals gain increased confidence in their decision-making process or require less information from their environment to make choices. Conversely, the more gradual decay of the LRA group could be attributed to their continuous evaluation and adaptation of strategies, necessitating a consistently prolonged decision-making time. In summary, this finding highlights the significance of considering the interplay between time and adaptation when examining human behavior within the context of serious games and decision-making processes.

\begin{table}[!hbp]%[H]
    \centering
    \begin{tabular}{|l|l|l|l|l|l|l|}
        \hline
        Group & mean [ms] & std [ms] & a & k & n\\ \hline
        RT  & -555.87 & 235.96 & 4.2603 & 0.3321 & 337 \\
        LRT & 686.51 & 285.90 & 4.3152 & 0.3282 & 279 \\
        LRA & 274.00 & 625.77 & 4.2547 & 0.2810 & 228 \\
        RA  & -326.90 & 377.50 & 4.3068 & 0.3650 & 251 \\ \hline
    \end{tabular}
    \caption{The first two columns show the mean and standard deviation of the median values by treatment type of the difference between each player and the average time of the entire population aggregated by behavioral group. The third and fourth columns show the parameter values of the power-law decay of the median time to complete a round by group. $f(t) = at^{-k}$. The group that reduces their time the fastest is the RA, with a ~30\% faster decay than the slowest group, the LRA. The last column (n) shows the size of each group.}
    \label{table:time}
\end{table}

\subsection*{Sensitivity to Contextual Information}\label{contextual}

The impact of communication~\cite{Granell2013} and response to incomplete information~\cite{incomplete} on immunization strategies has been demonstrated in previous studies. Our research extends this understanding by examining individual player behavior in response to specific treatment settings, such as different risk communication strategies, uncertainty levels of the nature of the spreading disease, and social uncertainties and the interplay of all of those variables. Studying how people react when facing this diverse set of certainty conditions is important as it provides critical insights into human decision-making processes, which is fundamental to predicting behavioral responses, essential for the modeling and forecasting of disease spread and thus, its containment~\cite{bavel2020using}.

Each treatment presented to the player is randomly selected at each round, consisting of a combination of two given levels of five variables. These variables include the average biosecurity level of the other 49 facilities, biosecurity uncertainty (which varies the number of farms whose information is shown or hidden to the players), contagion rate (representing the infection probability of a farm, which is also a function of the proximity to an infected farm), disease uncertainty (similar to biosecurity uncertainty, but for infection status), and messaging of the contagion rate level, which could be presented either as a gauge or a numeric value.

To analyze the impact of these treatment settings on player behavior, we compute the mean values of each treatment type and grouped them by their treatment type level, resulting in 16 observations for each group. By comparing the differences in player behavior across various treatment settings, our study aims to provide insights into how individuals adapt their strategies in response to changes in information and messaging, ultimately enhancing our understanding of effective immunization strategies in real-world contexts.

% For this we computed the median values of each treatment type an grouped them by their treatment type level which then will have 16 observations each.

% Fig 4
\begin{figure}[!hbp]
    \includegraphics[width=\textwidth]{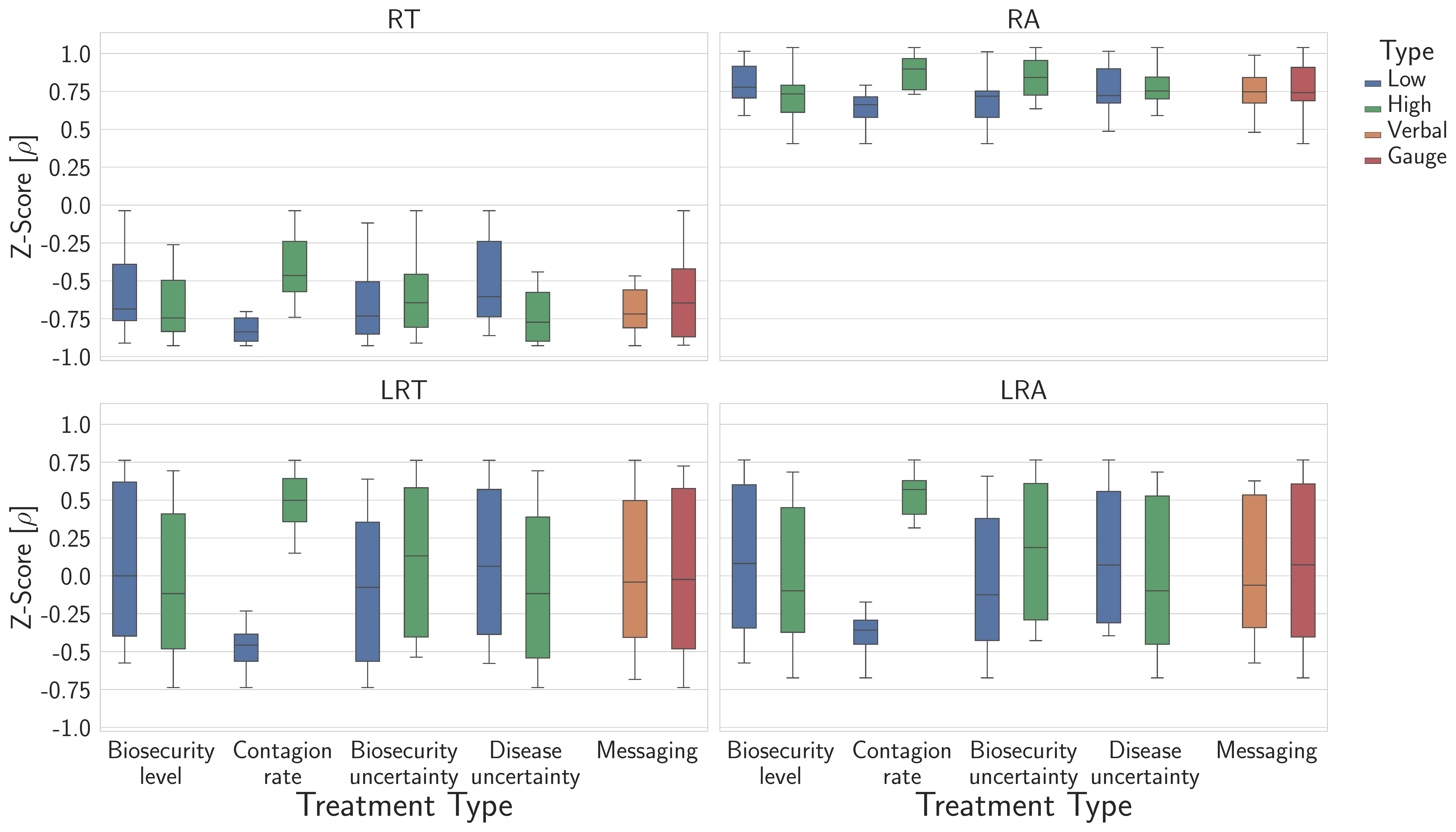}
\caption{Segmenting by the levels of each treatment type of the 32 rounds, we find that the contagion rate has the more drastic and the only statistical significant effect on contextualizing behaviors and the propensity for behavior change. }
\label{fig:treatmentypes}
\end{figure}

The contagion rate was the condition leading to the largest change in the $\rho$ value between treatments, as depicted in Figure~\ref{fig:treatmentypes}. In fact, for an $\alpha = 0.05$ it represents the only significant factor (Mann-Whitney, $U = 0$, $p = 2e-06$, two-sided). The other factors failed to exhibit significant differences (Mann-Whitney, $U \geq 87$, $p \geq 1.26e-01$, two-sided), implying that the disease's infectiousness was the most critical information for players when making decisions in the given environment, and providing evidence that players do indeed adapt their behavior based on contextual information.

To further investigate whether there was a strong signal for behavioral change conditioned to a fixed contagion rate value, we examined both low and high contagion rate cases. For the low contagion rate scenario, the most considerable change was influenced by biosecurity uncertainty (Mann-Whitney, $U = 11.0$, $p = 0.0313$, two-sided), which also held true for the high contagion rate (Mann-Whitney, $U = 12.0$, $p = 0.0405$, two-sided). Disease uncertainty exhibited a significant difference only for the low contagion rate (Mann-Whitney, $U = 52.0$, $p = 0.040$, two-sided), while the remaining variables did not display significant differences under either condition (Mann-Whitney, $p \geq 6.12-02$, two-sided).

These observations imply that once players have established the transmission rate, the subsequent most crucial pieces of information are the knowledge of which individuals are protected and which are not, followed by their infection status. This insight highlights the critical role that accurate and comprehensive information plays in shaping decision-making strategies, particularly in contexts pertaining to infectious diseases and biosecurity.

Knowledge of how individuals adapt their strategies in response to changes in different conditions can be of great help in the design and implementation of global and targeted interventions. Therefore, these findings could guide the creation of communication strategies and educational programs, ensuring that information presented to the public during a disease outbreak is effective in promoting adaptive behaviors and optimal decision-making.
% This means that when the players already established the transmission rate the next important piece of information is knowing who is protected and who is not followed by their infection status.

\begin{figure}[!hbp]
    \centering
    \includegraphics[width=\textwidth]{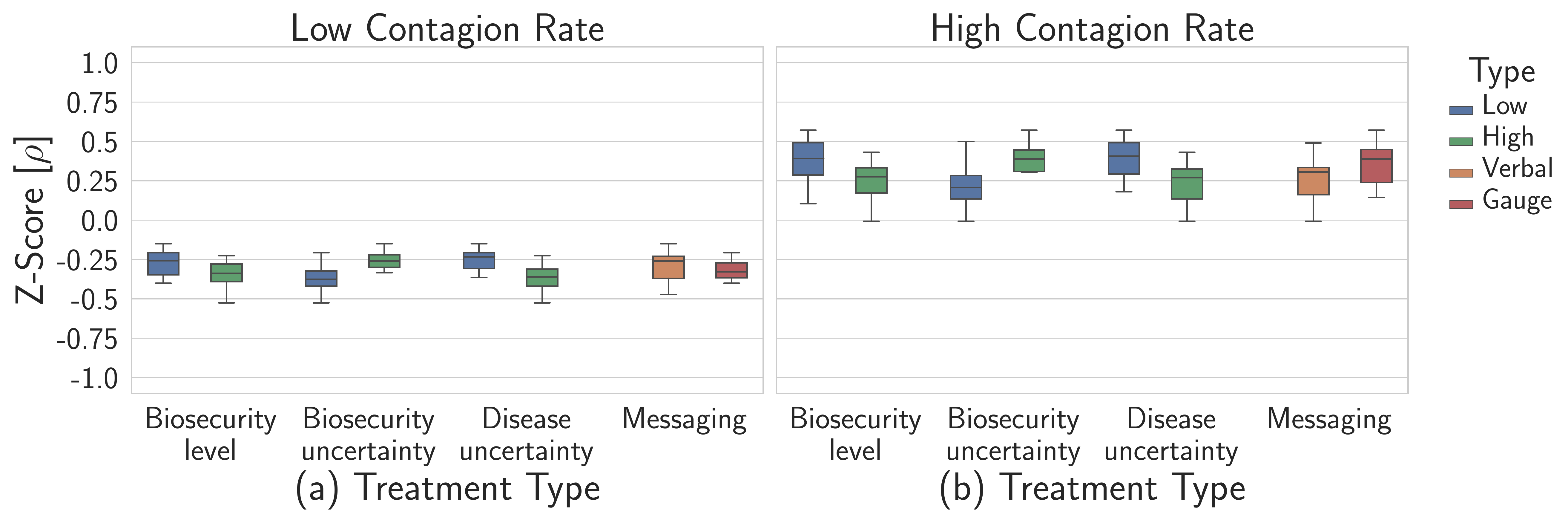}
    \caption{Median effect of treatment type, conditioned on the level of Contagion Rate. Interestingly, when controlling for the contagion rate, the difference in risk aversion due to uncertainty in biosecurity becomes significant at both low and high contagion rates. This implies that once players have established the risk associated with the contagion rate, they begin to assess the risk from the biosecurity measures taken by their neighbors. A particularly interesting observation, although not statistically significant, is the reversal in the effectiveness of verbal versus gauge messaging when controlling for contagion rate. Verbal messaging is more effective at low contagion rates, while gauge messaging is more effective at high contagion rates.}
    \label{fig:lowhigh}
\end{figure}

\section*{Discussion}

After analyzing nearly $32,000$ game rounds from $1,095$ players of a disease-spreading simulation game, we have identified at least four major groups of behavioral responses towards the phenomenon. Our findings provide evidence that some individuals do not react to information treatments or personal experience, and instead, demonstrate steady decision-making based solely on their established strategy. In contrast, another portion of the population is more responsive, adapting to their environment in two distinct ways: either learning to protect themselves or neglecting investment in protection. This experimental simulation can be employed to identify individuals with specific risk profiles and those who are prone to behavioral changes, enabling targeted interventions that encourage safer attitudes towards infection risk.

Furthermore, we have shown that adaptive groups tend to spend significantly more time making decisions compared to non-learning ones. We believe that this result could be utilized to design information treatments that promote analytical thinking within the population, ultimately increasing the adoption of biosecurity measures. This player segmentation was achieved through a fully automated method, shedding light on the potential for data-driven interventions.

While epidemiological models have incorporated a social parameter to account for knowledge contagion, they have thus far largely neglected to include vaccine-hesitant groups~\cite{Siddiqui2013}, whose numbers are unfortunately on the rise. To improve forecasts and public policies, it is crucial to account for these groups in our models, as they represent a segment of the population that may not adopt necessary prophylactic measures and consequently pose an inherent risk.

Our findings suggest that individuals primarily react to information regarding a disease's contagiousness when altering their attitude towards it. However, on a secondary level, they are also influenced by the protective measures taken by those around them. The significance of these results lies in their implications for encouraging individuals to invest in biosecurity measures, highlighting the importance of delivering accurate information through a well-crafted risk communication strategy~\cite{SALMON2015D66}.

These results are especially relevant in light of the recent COVID-19 pandemic. Our findings reveal that individuals respond strongly to information about a disease's contagiousness, which suggests that clear, transparent communication about disease transmissibility is crucial in shaping public attitudes and behavior during a disease outbreak.
Additionally, our study shows that information about the protective measures taken by others significantly influences individual behavior. This highlights the potential for leveraging social norms and observational learning to encourage adherence to biosecurity measures. During the COVID-19 pandemic, for instance, knowledge about others' adherence to measures such as mask-wearing or vaccination could influence an individual's own decision to adopt these protective behaviors~\cite{bavel2020using,muto2020japanese,latkin2021trust}.

Although we found correlations between behavioral change and multiple variables, such as the number of infections incurred, the precise factors leading an individual to adopt such changes remain unclear. Identifying these factors would be invaluable for utilizing simulations as interventions. Additionally, the time frame for the lasting effects of these games is unknown, which is a critical aspect for the effective and timely application of tools like these.

To address these questions, future research should include varying time intervals for follow-ups and multiple iterations of the experiment. This approach would allow for the assessment of social awareness by comparing participants' responses to information about previous players on the screen. It would also be important to test the impact of targeted interventions during time of game-play such as changing messaging strategies or informational pop-ups.

\section*{Methods} \label{methods}
We used an epidemic spread game simulation that takes place in a porcine farm network~\cite{merrill2019decision}. This game has been previously studied and discussed in the work by Clark et al.~\cite{clark2019using}, where they performed two experiments; in the first they recruited 50 players with a professional background in the pork farming industry at the 2018 American World Pork Expo and compared their result to a similar general audience in the Amazon Mechanical Turk (AMT) platform, finding that there was no significant difference between groups, and claim that the results from a larger study of non-professionals can be extrapolated to the target population, i.e. farming personnel. The second experiment was a total deployment on AMT with a participation of over 1000 players. The data used in this study is comprised by their results.\\

\subsection*{Game}\label{gamemethods}

The game simulation consists of 32 rounds with a maximum of six turns each, where 50 porcine farms are randomly placed in a constrained rectangular area. The management of one of these facilities is assigned to the player, while the other 49 are computer-controlled. At the beginning of a round, one of the computer-controlled facilities has an \textit{Infected} status, which can propagate through the rest of the facilities. The goal is to prevent the player's facility from becoming infected.

Each round provides the player with \$25,000 simulation dollars, and at each turn, they have the opportunity to increase their farm's biosecurity level in three incremental steps (\textit{None}, \textit{Low}, \textit{Medium}, \textit{High}) at a cost of \$1,000 simulation dollars or to remain at the same level. An increase in biosecurity level decreases the probability of infection. Levels and scores are not carried over to subsequent rounds, so each round starts with a biosecurity level of \textit{None} and \$25,000 simulation dollars, regardless of previous actions. The sum of all round scores determines the final score.

Computer-controlled facilities have two basic intrinsic properties: biosecurity level and infection status. Both properties can be hidden from the player, and the number of facilities with information displayed depends on the treatment being played. The facilities not controlled by the player are randomly initialized with a given biosecurity level, and only one of them has an \textit{infected} status. The biosecurity level remains constant throughout the round. The treatments consist of combinations of two levels/types of game variables: Average Biosecurity Level, Biosecurity Uncertainty, Contagion Rate, Disease Uncertainty, and Messaging. All of these have \textit{Low} and \textit{High} settings, except for Messaging, which can be either \textit{Verbal} or a \textit{Gauge}. Average Biosecurity Level controls the mean biosecurity level of the autonomous farms, Biosecurity Uncertainty determines the number of farms with hidden biosecurity levels, Contagion Rate represents the probability of an infected facility spreading the disease to its neighbors (which also scales with the Euclidean distance), Disease Uncertainty is similar to Biosecurity Uncertainty but regarding the biosecurity level, and Messaging communicates the infection probability to the player. These treatments yield $2^5=32$ different scenarios, with the presentation order to the players being entirely random.

If a player incurs an infection, \$25,000 is deducted from their accumulated simulation earnings, and the round ends immediately. At the end of the game, players are paid in U.S. dollars at a rate of \$50,000 simulation dollars = \$1 USD to encourage engagement.

\subsection*{Decision Space Embedding}
We are interested in the investment rate in the rounds and how it evolve during game. For this reason we formulated a round risk aversion metric $\rho_r = \frac{\beta(bl)_r}{\min(d_3,\tau_r)}$ where $\beta_r(bl)$ is a numerical mapping for the final biosecurity level acquired in the round given the following ordinal equivalence: $\beta(\text{None})=0$, $\beta(\text{Low})=1$, $\beta(\text{Medium})=2$, $\beta(\text{High})=3$. The term $\min(d_H,\tau_r)$ represents the number of opportunities during which players had the chance to invest in biosecurity, with $d_H$ beign the number of turns required to reach a \textit{High} level and $\tau_r$ the duration in turns of the round $r$ as the round is automatically terminated if an infection happens, $6$ is the maximum value for $\min(d_H,\tau_r)$ and 3 is the minimum. This metric captures the proportion of investment relative to the opportunity to invest, such that a player with $\rho_r =1$ invests in more biosecurity at every available opportunity, while a player with $\rho_r =0$ never invests.

\subsection*{Clustering Users}
\subsubsection*{Isomap}
The Isomap algorithm~\cite{balasubramanian2002isomap} is a dimensionality reduction technique that belongs to the family of manifold learning algorithms. By applying Isomap to the data, the inherent structure is preserved, which facilitates the analysis and understanding of complex patterns. The algorithm consists of the following steps:
\begin{enumerate}
\item Nearest neighbor search: Find the nearest $k$ vectors to each representation of a player's game using Euclidean distance. We then create a graph connecting each point to it's nearest $k$ neighbors.
\item Geodesic distance: Create a matrix representation of the graph using the geodesic distance between each pair of vectors. This is done using the Dijkstra's algorithm~\cite{dijkstra1959note}.
\item Multidimensional scaling (MDS)~\cite{torgerson1952multidimensional}: Using the given matrix, find the $n$ smallest eigenvalues and their corresponding eigenvectors. These eigenvectors will be the new dimensions, each entry of these eigenvectors represents a player.
\end{enumerate}

We applied the Isomap algorithm to project the 32-dimensional decision data, representing each round, onto a 2D space using $k = 20$ nearest neighbors and kept the first 2 eigenvectors to create the 2D representation.

\subsubsection*{K-means}
Given the 2D representation of the data we used the K-means algorithm to cluster users with similar trajectories.
The K-means algorithm works as follow:
\begin{enumerate}
   \item Initialize $K$ centroids randomly within the data points. 
   \item Measure the Euclidean distance from all the data points to all of the centroids.
   \item Assign each point to the closest centroid.
   \item For each centroid group, compute the mean of the coordinates of all the points in the group.
   \item Repeat 3,4 and 5 until convergence.
\end{enumerate}

In order to find the optimal number of partitions we use the distortion score. The distortion score is calculated as the sum of the squared distances between each data point in a cluster and its correspondent centroid. We then plot the distortions scores against the $K$ values and identify the point at which the distortion score "slows down" its growth ("elbow"). This elbow point serves as the optimal number of clusters for our analysis.

\bibliography{attitudes}

 \section*{Acknowledgements}
This work was supported by the USDA National Institute of Food and Agriculture, under award number 2015-69004-23273. The contents are solely the responsibility of the authors and do not necessarily represent the official views of the USDA or NIFA.

% Acknowledgements should be brief, and should not include thanks to anonymous referees and editors, or effusive comments. Grant or contribution numbers may be acknowledged.

\section*{Author contributions statement}
OL conceived the experiment, analysed the data, and wrote the initial manuscript. NC supervised the analysis and contributed to the initial manuscript draft. Data acquisition and game experiments were conducted by SM and EC. SM, GB, EC, TL, TS, CK, AZ, and JS contributed to the writing, review and editing of the manuscript.

% \section*{Additional information}

% To include, in this order: \textbf{Accession codes} (where applicable); \textbf{Competing interests} (mandatory statement). 

% The corresponding author is responsible for submitting a \href{http://www.nature.com/srep/policies/index.html#competing}{competing interests statement} on behalf of all authors of the paper. This statement must be included in the submitted article file.

\end{document}